\documentclass[paper,twocolumn,showpacs,superscriptaddress,nofootinbib,preprintnumbers,floats,aps,pra,10pt]{revtex4-1}
\usepackage{graphicx}
\usepackage{dcolumn}
\usepackage{amsmath}
\usepackage{amsfonts}
\usepackage{amssymb}
\usepackage{color}
\usepackage{float}
\usepackage{pstricks}
\usepackage{natbib}
\usepackage{array}
\usepackage{multirow}
\usepackage{ulem}
\providecommand{\U}[1]{\protect\rule{.1in}{.1in}}

\DeclareMathOperator\arctanh{arctanh}

\begin{document}

\title{Study of a viscous $\Lambda$WDM model: Near equilibrium condition, entropy production, and cosmological constraints}
\author{Norman Cruz}
\altaffiliation{norman.cruz@usach.cl}
\affiliation{Departamento de F\'isica, Universidad de Santiago de Chile, \\
Avenida Ecuador 3493, Santiago, Chile}
\affiliation{Center for Interdisciplinary Research in Astrophysics and Space Exploration (CIRAS),Universidad de Santiago de Chile, Avenida Libertador Bernardo O'Higgins 3363, Estación Central, Chile}
\author{Esteban Gonz\'alez}
\altaffiliation{esteban.gonzalez@uac.cl}
\affiliation{Direcci\'on de Investigaci\'on y Postgrado, Universidad de Aconcagua, Pedro de Villagra 2265, Vitacura, 7630367 Santiago, Chile}
\author{Jose Jovel}
\altaffiliation{jose.jovel@usach.cl}
\affiliation{Departamento de F\'isica, Universidad de Santiago de Chile, \\
Avenida Ecuador 3493, Santiago, Chile}

\date{\today}

\begin{abstract}
\begin{center}
\textbf{Abstract:}    
\end{center}

Extensions to a $\Lambda$DM model have been explored to face current tensions that occur within its framework, which encompasses broadening the nature of the dark matter (DM) component to include warmness and a non-perfect fluid description. In this paper, we investigated the late-time cosmological evolution of an exact solution recently found in [N. Cruz, E. Gonz\'alez, and J. Jovel, Phys. Rev. D \textbf{105}, 024047 (2022)], which describe a viscous warm $\Lambda$DM model ($\Lambda$WDM) with a DM component that obeys a polytropic equation of state (EoS), which experience dissipative effects with a bulk viscosity proportional to its energy density, with proportionality constant $\xi_{0}$. This solution has the particularity of having a very similar behavior to the $\Lambda$CDM model for small values of $\xi_{0}$, evolving also to a de Sitter type expansion in the very far future. We explore firstly the thermodynamic consistences of this solution in the framework of the Eckart's theory of non-perfect fluids, focusing on the fulfillment of the two following conditions: (i) the near equilibrium condition and (ii) the positiveness of the entropy production. We explore the range of parameters of the model that allow to fulfilling these two conditions at the same time, finding that a viscous WDM component is compatible with both ones, being in this sense, a viable model from the thermodynamic point of view. Besides, we constraint the free parameters of the model with the observational data coming form supernovae Ia (SNe Ia) and the observational Hubble parameter data (OHD), using these thermodynamics analysis to define the best priors for the cosmological parameters related to the warmness and the dissipation of the DM, showing that this viscous $\Lambda$WDM model can describe the combined  SNe Ia + OHD data in the same way as the $\Lambda$CDM model. The cosmological constraint at $3\sigma$ CL give us an upper limit on the bulk viscous constant of the order $\xi_{0}\sim10^{6}Pa\times s$, which is in agreement with some previous investigations. Our results support that the inclusion of a dissipative WDM, as an extension of the standard cosmological model, leads to a both thermodynamically consistent and properly fitted cosmological evolution.

\vspace{0.5cm}
\end{abstract}
 \maketitle


\section{Introduction}

In the current cosmology, the observational evidence suggests that almost the total energy density of the Universe is compound by the dark sector, roughly classified into 30\% of dark matter (DM) and 70\% of dark energy (DE) \cite{Planck2018,WMAP2013,BAO2017,OHD2021}. The DM is responsible for the structure formation in the Universe while the DE is responsible for the recent accelerating expansion of the Universe \cite{sPerlmutter}. The most simple model that includes these components and fits very well the cosmological data is the $\Lambda$CDM model \cite{Planck2018,WMAP2013}, where the DE is modeled by a positive cosmological constant (CC) $\Lambda$, and the DM is described as a pressureless fluid known as cold DM (CDM). However, this model is not absent of problems from the theoretical and observational point of view. For example, the CC problem, where the value of the CC differs from theoretical field estimations in 60-120 orders of magnitude than the observed value \cite{CPWeinberg,Lambda,problemadecoincidenciaylambda}. Also, measurements of the Hubble parameter at the current time, $H_{0}$, presents a discrepancy of $4.4\sigma$ between the measurements obtained from Planck CMB and the locally measurements obtained by A. G. Riess \textit{et al.} \cite{Riess:2019cxk}. Other tensions are the measurements of $\sigma_{8}-\Omega_{m}$ (where $\sigma_{8}$ is the r.m.s. fluctuations of perturbations at $8h^{-1}Mpc$ scale) coming from large scale structure (LSS) observations and the extrapolated from Planck CMB (dependent on the $\Lambda$CDM model) \cite{remedyforplankanlssdata,sigma8}, and the results from the experiment to detect the global EoR signature (EDGES), which detect an excess of radiation in the reionization epoch that is not predicted by the $\Lambda$CDM model, specifically at z$\approx 17$ \cite{linia21cm}.


In order to try to overcome some of these problems, dissipative effects can be considered as a more realistic way of treating cosmic fluids. In a homogeneous and isotropic Universe, the dissipative process is usually characterized by a bulk viscosity. In this sense, some authors have considered the non-inclusion of the CC in order to alleviate the CC problem, explaining the late time acceleration behavior of the Universe through a dissipative viscous fluid \cite{paperprofeAccelerated,Bulk1,Bulk2,Bulk3,Bulk4,Bulk5,Bulk6,Bulk7,Bulk8,Exact,Testing,Almada1}, as a natural choice since the effect of the bulk viscosity is to produce a negative pressure that leads to an acceleration in the universe expansion \cite{Oditsov1,Oditsov2,Almadaconstraints}. Also in \cite{tensionH0,BulktensionH} the authors discuss the $H_{0}$ tension problem in the context of dissipative fluids as a good chance to construct new cosmological models with non-ideal fluids. Even more, the $\sigma_{8}-\Omega_{m}$ tension can be alleviated if one assumes a small amount of viscosity in the DM component \cite{remedyforplankanlssdata}, as well as the explanation of the excess of radiation predicted by the EDGES experiment \cite{linea21cmH}.

On the other hand, bulk viscosity seems to be significant in the cosmic evolution. For example, many observational properties of disk galaxies can be represented by a dissipative DM component \cite{foot2015dissipative,foot2016solving}. For neutralino CDM, the bulk viscous pressure is present in the CDM fluid through the energy transferred from the CDM fluid to the radiation fluid \cite{hofmann2001damping}. Some authors propose that bulk viscosity can produce different cooling rates of the components of the cosmic fluid \cite{cooling1,cooling2,cooling3}, or may be the result of non-conserving particle interactions \cite{Cosmologycreation}. Even more, from Landau and Lifshitz \cite{LandauandLifshitz}, the bulk viscosity can be interpreted from the macroscopic point of view as the existence of slow processes to restore the equilibrium state. At perturbative level, viscous fluid dynamics provides also a simple and accurate framework with the purpose of extending the description into the nonlinear regime \cite{blas2015large}. Following this line, the bulk viscosity $\xi$ depends, particularly, on the temperature and pressure of the dissipative fluid \cite{librocaro}. Therefore, a natural election for the bulk viscosity of the dissipative fluid is to consider a dependency proportional to the power of their energy density $\xi=\xi_{0}\rho^{m}$, where $\xi_{0}>0$ is a bulk viscous constant; election that has been widely investigated in the literature \cite{Big.Bang,rho1,Brevik,Analysing,valorxi1,valorxi2}. Since the nature of the DM it is unknown up to date, and dissipative effect can not be discarded \cite{Oditsov3}, it is of physical interest to explore how a bulk viscous DM behaves in the $\Lambda$CDM model.


Another important possible extension of the standard cosmological model has also been investigated in the last decades. It is well known that a CDM is capable to explains the observed structure very well above $\thicksim 1 Mpc$, while it has issues explaining small-scale structure observations \cite{WDMgalaxias2,WDMgalaxies} such as the missing satellite problem \cite{ProblemSat1,problemasatelite2}, that refers to the discrepancy of about 10 times more dwarf galaxies between the values obtained by the numerical simulations based on $\Lambda$CDM model and the observed ones in cluster of galaxies. In this sense, a warm DM (WDM) can potentially be a good candidate to explain small-scale structure observations that currently represent a challenge for the CDM model. Many studies of the number of satellites in the Milky Way or small halos with dwarf galaxies appear to be in better agreement with the observations for a WDM than they are for a CDM \cite{wdmobs,wdm3,WDMgalaxies,WDMgalaxi3}. 
One of the well motivated WDM hypothesis implies an extension of the standard model of particle physics by three sterile (right-handed, gauge singlet) neutrinos \cite{wdm0,sneutrino,sneutrin1}, produced via mixing with active neutrinos in the early Universe \cite{wdm0,creacionSneutrino,creacionSneutrino1,creacionSneutrino2,creacionSneutrino3,creacionSneutrino4}. On the other hand, from the perturbation point of view, the no-linear effects make the power-spectum of the WDM look very much like a CDM, and LSS such as filaments, sheets, and large void suggest that a WDM reproduces well the observed ones \cite{Perturbaciones1}; being the WDM an interesting alternative from the point of view of cosmology and particle physics. 

All the discrepancies mentioned above implies extensions of the $\Lambda$CDM model like the considerations of dissipative effects in a WDM component, which leads to taken into account a relativistic thermodynamic theory of non-perfect fluids out of equilibrium. Eckart was the first to develop such theory \cite{Eckart}, with a similar model proposed by Landau and Lifshitz \cite{LandauandLifshitz}. However, it was later shown that Eckart's theory was a non-causal theory \cite{NocausalEckart,Muller}. A causal theory was proposed by Israel and Stewart (IS) \cite{I.S.1,I.S.2}, which is reduced to Eckart's theory when the relaxation time for the bulk viscous effects are negligible \cite{Dissipativecosmology}. Since the IS theory presents a much greater mathematical difficulty than the Eckart's theory, this last one is considered as a first approximation in order to study viscous cosmology \cite{Big.Bang,AlmadaEckartbulktest,bulkyaceleracion,chinos,articuloprofephantom,Brevik}. Following this line, the authors in \cite{articulobueno} study a universe filled by two fluids in the framework of Eckart's theory, a perfect fluid as DE mimicking the dynamics of the CC, and a bulk viscous DM, finding a good agreement at $3\sigma$ CL with the observational background data coming from the observational Hubble parameter (OHD), type Ia supernovae (SNe Ia), and strong lensing systems (SLS). It is important to mention that, in these theories, the bulk viscous pressure $\Pi$ has to be lower than the equilibrium pressure $p$ of the dissipative fluid, i. e.,
\begin{equation}\label{nearequilibrium}
l=\left|\frac{\Pi}{p}\right|\ll 1,
\end{equation}
which it is known as \textit{the near equilibrium condition}, and represents the assumption that the fluid is close to thermodynamic equilibrium. 

According to Maartens, in the context of dissipate inflation \cite{Dissipativecosmology}, the condition to have an accelerated expansion due only to the negativeness of the viscous pressure $\Pi$ in the Eckart's and IS theories enters into direct contradiction with the near equilibrium condition given by Eq. \eqref{nearequilibrium}. In this sense, as has been proposed in \cite{Analysing,articulobueno}, if a positive CC is considered in these theories, then the near equilibrium condition could be preserved in some regime. In addition,  it was shown by J. Hua and H. Hu \cite{chinos} that a dissipative DM in Eckart's theory with CC has a significantly better fit with the cosmological data than the $\Lambda$CDM model, which indicates that this model is competitive  to fit the combined SNe Ia + CMB + BAO + OHD data. Nevertheless, the inclusion of the CC implies to abandon the idea of unified DM models with dissipation, whose advantage is to avoid the CC problem, but that leads to reinforce the proposal of extending the standard model, keeping a DE component modeled by a CC. 
Another important point of the near equilibrium condition given by the Eq. \eqref{nearequilibrium} is that we need a non-zero equilibrium pressure for the dissipative fluid, discarding the possibility of a CDM. In this sense, a relativistic approach of dissipative fluids is consistent with a WDM component \cite{WDMgalaxias2,wdm0,wdm1,wdm2,wdm,wdm3,wdm4} in order to satisfy the near equilibrium condition.

It is important to mention that, for cosmologies with perfect fluids, there is no entropy production because these fluids are in equilibrium and their thermodynamics are reversible. But, for cosmologies with non-perfect fluid, where irreversible process exists, there is a positive entropy production during the cosmic evolution \cite{Tamayo,fullsolution,Exact,entropiaprofe,Mar2}. The near equilibrium condition and entropy production has been previously discussed in the literature. The near equilibrium condition was studied, for example, in \cite{fullsolution} for the IS theory with gravitational constant $G$ and $\Lambda$ that vary over time; while in \cite{DissipativeBoltzmann} it was studied in the Eckart's and IS theories for the case of a dissipative Boltzmann gas and without the inclusion of a CC. The entropy production was studied in \cite{Tamayo} in the Eckart's and IS theories for a dissipative DE; while in \cite{Exact} the authors study the entropy production in the full IS theory with a matter content represented by one dissipative fluid component, and the kinematics and thermodynamics properties of the solutions are discussed (the entropy production in cosmological viscous fluids has been more widely studied, and more references can be found in \cite{producciondeentropia,Titus,produccionentropia2,produccionentropia3,produccionentropia4,produccionentropia5}).

The aim of this paper is to explore the thermodynamic consistency in the description of a dissipative WDM component, not enough investigated up to date, and if the constraints from the present cosmological data on the model that we propose are compatible with the consistency criteria found. Our model is described by an analytical solution obtained in \cite{primerarticulo}, for a flat Friedman-Lemaître-Robertson-Walker (FLRW) universe, dominated by a dissipative DM modeled by the barotropic EoS $p=(\gamma-1)\rho$, where $\rho$ is the energy density of the dissipative DM and $\gamma$ is known as barotropic index, and a DE modeled by the CC, in the framework of the Eckart's theory. This solution was obtained using the expression $\xi=\xi_{0}\rho^{m}$ for the bulk viscosity with the particular choice of $m=1$, and was studied in the context of the late and early-times singularities. Although this solution was found for a very particular election of $m$, has the important characteristics that, for a positive CC, behaves very similarly to the $\Lambda$CDM model for all the cosmic time when $\xi_{0}\to 0$, without singularity towards the past in an asymptotic behavior known as ``soft-Big Bang'' \cite{Softbang1,SoftBang2}, and with an asymptotic de Sitter expansion towards the future. This last behavior is of interest because the solution tends to the de Sitter expansion regardless of the value of $\xi_{0}$ and $\gamma$, as long as $\xi_{0}<\gamma/3H_{0}$, which is a feature not found for other elections of $s$. Therefore, we focus our study to the late-time behavior of this solution assuming that $\gamma\neq 1$ but close to $1$, which represents a dissipative $\Lambda$WDM model with the same asymptotic late-time behavior that the $\Lambda$CDM model. In particular, we study the near equilibrium condition and the positiveness of entropy production of this solution to find the constraints that these criteria impose on the model's free parameters. We focus on the possibility to have a range of them satisfying all of these conditions, and compare it with the best fit values obtained from the cosmological constraint with the SNe Ia + OHD data. This study leads to obtaining some important clues about the physical behavior of the analytical solution, which represents a particular extension of the standard cosmological model. In this model, we have two more free parameters than the $\Lambda$CDM model, namely, $\gamma$ and $\xi_{0}$, but the gain lies in a more complete description of the nature of the DM component, suggested  by previous investigation made in the context to alleviate tensions in the standard cosmological model. Despite the fact that we are not facing in this work none of the mentioned tensions, our primary intention is to explore if the two extension made to the standard model present also a consistent relativistic fluid description and not only a well suitable fit with the cosmological data.

The outline of this paper is as follows: In Sec. \ref{seccion2} we summarize a solution that was found in \cite{primerarticulo} which represent the model of our study. In Sec. \ref{tresproblemas} we present general results about the near equilibrium condition and the entropy production of the viscous fluid. In Sec. \ref{exactlate} we study the solution at late-times, where in Sec. \ref{seccion4A} we study the fulfillment of the near equilibrium condition, while in Sec. \ref{seccion4C} we study the entropy production of the dissipative fluid present in the model. In Sec \ref{constraints} we constrain the free parameters of our model with the SNe Ia and OHD data. In Sec. \ref{results} we discuss this
results, comparing them with the $\Lambda$CDM model, and we study the completeness of both, the near equilibrium condition, and entropy production for the actual data. In addition, we find a upper limit for the present value of the bulk viscous constant. Finally, In Sec. \ref{seccionfinal} we present some conclusions and final discussions. $8\pi G=c=1$ units will be used in this work.

\section{Exact analytical solution in Eckart's theory with CC}\label{seccion2}

In this section, we briefly resume a de Sitter-like solution and an analytical solution found in \cite{primerarticulo}. For a flat FLRW universe composed with a dissipative DM ruled by the barotropic EoS $p=(\gamma-1)\rho$, with a bulk viscosity of the form $\xi=\xi_{0}\rho^{m}$, and DE given by the CC, the field equations, in the framework of Eckart's theory, are given by \cite{Eckart,primerarticulo}
\begin{equation}\label{tt}
H^2=\Big(\frac{\dot{a}}{a}\Big)^{2} =\frac{\rho}{3}+\frac{\Lambda}{3},
\end{equation}
\begin{equation}\label{rr1}
\frac{\ddot{a}}{a}=\dot{H}+H^2=-\frac{1}{6} \left(\rho+3P_{eff}\right)+\frac{\Lambda}{3},
\end{equation}
\begin{equation}\label{ConsEq1}
\dot{\rho}+3H(\rho+P_{eff})=0,
\end{equation}
where
\begin{equation}\label{Peff}
P_{eff}=p+\Pi,
\end{equation}
\begin{equation}\label{Pi}
\Pi=-3H\xi.
\end{equation}
From this equations, it is possible to obtain a single evolution equation for the Hubble parameter $H=\dot{a}/a$, where $a$ is the scale factor and ``dot'' accounts for the derivative with respect to the cosmic time $t$, which is given by
\begin{equation}\label{Hpunto}
2\dot{H}+3\gamma H^{2}-3\xi_{0}H(3H^{2}-\Lambda)^{m}-\Lambda\gamma=0.
\end{equation}

We address the reader to see all the technical details in \cite{primerarticulo}, where, from Eq. \eqref{Hpunto}, a de Sitter-like and exact solutions for the cases of $m=0$ and $m=1$, with positive and negative CC, has been studied in the context of late and early-time singularities. The results were compared with the $\Lambda$CDM model and, for this purpose, the differential equation \eqref{Hpunto} is solved for $\xi_{0}=0$ with the initial conditions  $H(t=0)=H_{0}$ and $a(t=0)=1$, which leads to
\begin{equation}\label{Hestandar}
H(t)=\frac{H_{0}\sqrt{\Omega_{\Lambda_{0}}}\left(\left(\sqrt{\Omega_{\Lambda_{0}} }+1\right) e^{3 \gamma  H_{0} t \sqrt{\Omega_{\Lambda_{0}} }}-\sqrt{\Omega_{\Lambda_{0}} }+1\right)}{\left(\sqrt{\Omega_{\Lambda_{0}} }+1\right) e^{3\gamma H_{0} t \sqrt{\Omega_{\Lambda_{0}} }}+\sqrt{\Omega_{\Lambda_{0}} }-1},  
\end{equation}
\begin{equation}\label{aestantdar}
    a(t)=\left(\cosh\left(\frac{3\gamma\sqrt{\Omega_{\Lambda_{0}}}H_{0}t}{2}\right)+\frac{\sinh\left(\frac{3\gamma\sqrt{\Omega_{\Lambda_{0}}}H_{0}t}{2}\right)}{\sqrt{\Omega_{\Lambda_{0}}}}\right)^{\frac{2}{3\gamma}},
\end{equation}
where $\Omega_{\Lambda_{0}}=\Lambda/(3H^{2}_{0})$. Note that Eq.(\ref{Hestandar}) tends asymptotically at very late times ($t\to\infty$) to the de Sitter solution $H_{dS}=H_{0}\sqrt{\Omega_{\Lambda_{0}}}$. 

\begin{figure}[H]
\centering
\includegraphics[width=0.48\textwidth]{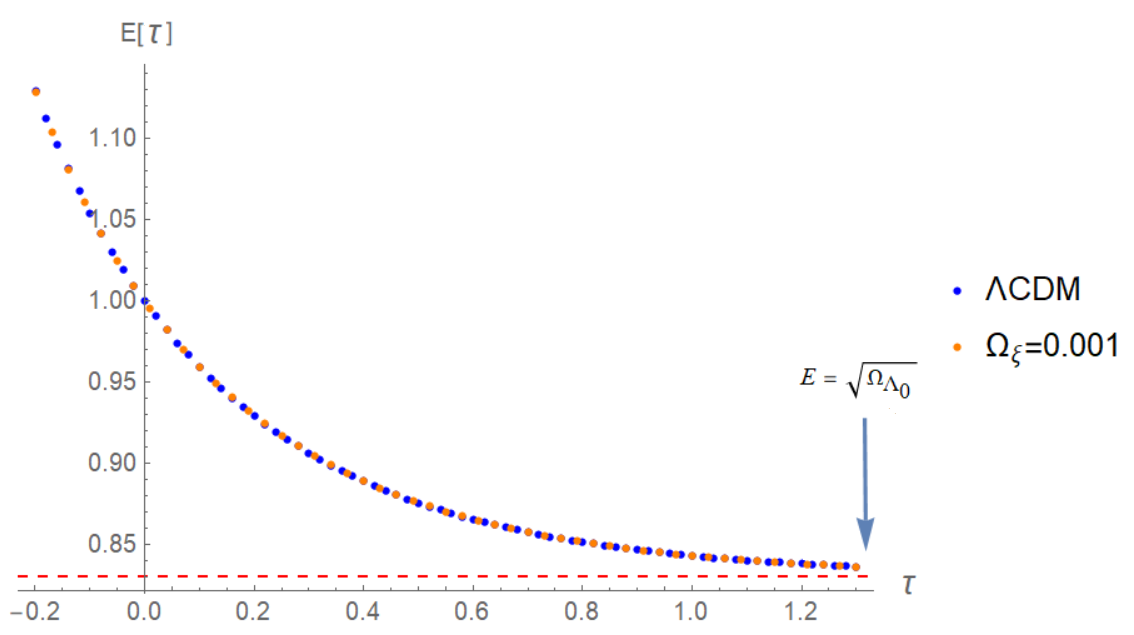}
\caption{Numerical behavior of $E(\tau)$ obtained from Eq. \eqref{Tm1} at late times, for $\Omega_{\Lambda_{0}}=0.69$, $\Omega_{\xi_{0}}=0.001$ and $\gamma=1.002$. For a comparison, we also plotted the $\Lambda$CDM model obtained from Eq. \eqref{Hestandar}.}
\label{figuraem1}
\end{figure}

We are particularly interested in the case of $m=1$ of Eq. \eqref{Hpunto} with a positive CC, where the following de Sitter-like solution ($\dot{H}=0$) are found
\begin{eqnarray}\label{HdeSitter}
 E_{dS}&=&\sqrt{\Omega_{\Lambda_{0}}},
\end{eqnarray}
with $E=H(t)/H_{0}$. It is important to note that Eq. \eqref{HdeSitter} is the usual de Sitter solution, written in its dimensionless form, and naturally appears in this dissipative scenario. On the other hand, the exact analytical solution ($\dot{H}\neq 0$) found, takes the following expression in terms of the dimensionless parameters
\begin{eqnarray}\label{Tm1}
&\tau=\frac{\Omega_{\xi_{0}} \sqrt{\Omega_{\Lambda_{0}}} \log \left(\frac{(1-\Omega_{\Lambda_{0}} ) (\gamma - E\Omega_{\xi_{0}} )^2}{\left(E^2-\Omega_{\Lambda_{0}} \right) (\gamma - \Omega_{\xi_{0}} )^2}\right)}{{3 \sqrt{\Omega_{\Lambda_{0}} } \left(\gamma^2-\Omega^{2}_{\xi_{0}} \Omega_{\Lambda_{0}} \right)}} \nonumber \\
& +\frac{\gamma  \log \left(\frac{\left(\sqrt{\Omega_{\Lambda_{0}} }-1\right) \left(\sqrt{\Omega_{\Lambda_{0}} }+E\right)}{\left(\sqrt{\Omega_{\Lambda_{0}} }+1\right) \left(\sqrt{\Omega_{\Lambda_{0}} }-E\right)}\right)}{{3 \sqrt{\Omega_{\Lambda_{0}} } \left(\gamma^2-\Omega^{2}_{\xi_{0}} \Omega_{\Lambda_{0}} \right)}},
\end{eqnarray}
obtained with the border condition $H(t=0)=H_{0}$, and where $\Omega_{\xi_{0}}=3\xi_{0}H_{0}$ and $\tau=tH_{0}$ are the dimensionless bulk viscous constant and cosmic time, respectively. The above solution is an implicit relation of $E(\tau)$. According to \cite{primerarticulo}, Eq. \eqref{Tm1} presents a future singularity in a finite time known as Big-Rip \cite{BigSmash(B.R),tiposdeBigRip,clasificacionBigRipdetallada,Brevik} when $\Omega_{\xi_{0}}>\gamma$. In this singularity we have an infinite $a$, $\rho$ and $p$, and, therefore, the Ricci scalar diverges. Also, is discussed that one interesting behavior of this solution can be obtained if we considered the opposite condition, i.e.,
\begin{equation}\label{condicionBigRip}
    \Omega_{\xi_{0}}<\gamma,
\end{equation}
which leads to a universe with a behavior very similar to the $\Lambda$CDM model, which coincide as $\Omega_{\xi_{0}}\to 0$, as can be seen in Fig. \ref{figuraem1}, where we have numerically found the behavior of $E$ as a function of $\tau$ from Eq. \eqref{Tm1}, taking into account the condition \eqref{condicionBigRip} with $\gamma=1.002$, $\Omega_{\xi_{0}}=0.001$, and $\Omega_{\Lambda_{0}}=0.69$. For a comparison, we also plotted the $\Lambda$CDM model. 

Note that, the solution \eqref{Tm1} tends asymptotically for $\tau\to\infty$ to the usual de Sitter solution \eqref{HdeSitter}, which can be seen in the Fig. \ref{figuraem1}. Therefore, for the condition given by Eq. \eqref{condicionBigRip} and for $\gamma\neq 1$ but close to $1$, solution \eqref{Tm1} represent a viscous $\Lambda$WDM model with a late-time behavior very similar to the $\Lambda$CDM model and with the same asymptotic de Sitter expansion.

\section{Near equilibrium condition and entropy production}\label{tresproblemas}

In what follows we found the main expressions in terms of $\gamma$, and the dimensionless quantities $E$,  $\Omega_{\xi_{0}}$ and $\Omega_{\Lambda_{0}}$, that arises from the near equilibrium condition  and the entropy production.

\subsection{Near equilibrium condition}
As it was previously discussed, in Eckart's theory it is necessary to fulfill the near equilibrium condition \eqref{nearequilibrium}. Following Maartens~\cite{Dissipativecosmology}, and according to the Eqs. \eqref{rr1} and \eqref{Peff}, we can write
\begin{equation}\label{rr}
\frac{\ddot{a}}{a}=-\frac{1}{6} \left[\rho+3\left(p+\Pi\right)\right]+\frac{\Lambda}{3}.
\end{equation}
From the above expression, the condition to have an accelerated expansion driven only by the negativeness of the bulk viscous pressure $\Pi$, imposing $\ddot{a}>0$ and taking $\Lambda =0$, is
\begin{equation} \label{Pieqa}
-\Pi >p+\frac{\rho}{3}.
\end{equation}
This last result implies that the viscous stress is greater than the equilibrium pressure $p$ of the fluid, i.e., the near equilibrium condition is not fulfilled because in order to obtain an accelerated expansion the fluid has to be far from equilibrium. This situation could be change if a positive CC is included \cite{Analysing,articulobueno}. In this case, the condition $\ddot{a}>0$ on Eq. \eqref{rr} leads to
\begin{equation}\label{Pieqa1}
-\Pi >\frac{-2\Lambda}{3}+p +\frac{\rho}{3},
\end{equation}
i.e., the near equilibrium condition could be fulfilled in some regime, because from Eq. \eqref{Pieqa1} the viscous stress not necessarily is greater than the equilibrium pressure $p$. The near equilibrium condition given by Eq. \eqref{nearequilibrium} can be rewritten in terms of the dimensionless parameters, using Eq. \eqref{Pi}, and the EoS of the DM component, obtaining
\begin{equation}\label{lhm1}
    l=\left|\frac{E(\tau)\Omega_{\xi_{0}}}{\gamma-1}\right|\ll 1.
\end{equation}    

From the above equation is clear to see that a CDM component with $\gamma=1$ is not compatible with the near equilibrium condition, and only for some kind of WDM with $\gamma >1$ this condition can be fulfilled. On the other hand, note that in the above expression, the solution given by $E(\tau)$ drives the behavior of $l$ as a function of the cosmic time $\tau$. In this sense, and since $E(\tau)$ is a decreasing function of time, the constraints on $\Omega_{\xi_{0}}$ are more restrictive as we look forward.

\subsection{Entropy production}\label{seccion2A}
The First law of thermodynamics is given by
\begin{eqnarray}\label{primeraley}
    TdS=dU+pdV, 
\end{eqnarray}
where $T$, $S$, $U$, $V$ are the temperature, entropy, internal energy, and the three dimensional volume of the cosmic fluid. The internal energy of the fluid and the physical three dimensional volume of the Universe are given respectively  by $U=\rho V$ and $V=V_{0}a^{3}$ (where $V_{0}$ is the volume at the present time). With these, we get from Eq. \eqref{primeraley} the Gibbs equation \cite{Mar2}
\begin{equation}\label{gibbspaso1}
    dS=-\left(\frac{\rho+p}{Tn^{2}}\right)dn+\frac{d\rho}{Tn},
\end{equation}
where $n=N/V$  is the number of particle density. The following integrability condition must hold on the thermodynamical variables $\rho$ and $n$
\begin{equation}
    \left[\frac{\partial}{\partial \rho}\left(\frac{\partial S}{\partial n}\right)_{\rho}\right]_{n}= \left[\frac{\partial}{\partial n}\left(\frac{\partial S}{\partial \rho}\right)_{n}\right]_{\rho},
\end{equation}
then, we considered the thermodynamic assumption in which the temperature is a function of the number of particles density  and the energy density, i.e., $T(n,\rho)$. With this, the above integrability condition become in \cite{Tamayo, Mar2}
\begin{eqnarray}\label{temperatura}
    n\frac{\partial T}{\partial n}+\left(\rho+p\right)\frac{\partial T}{\partial \rho}=T\frac{\partial p}{\partial \rho}.
\end{eqnarray}
We study the case of a perfect fluid and a viscous fluid separately in order to compare our result with the model without viscosity.

For a perfect fluid the particle 4-current is taken to be $n^{\alpha}_{;\alpha}=0$, where ``;" accounts for the covariant derivative, which together with the conservation equation, leads to the following expressions for the particle density and the energy density, respectively
\begin{eqnarray}\label{npunto}
    \dot{n}+3H n=\frac{\dot{N}}{N}&=&0,\\ \label{fluidoperfecto}
    \dot{\rho}+3H(\rho+p)&=&0.
\end{eqnarray}
Assuming that the the energy density depends on the temperature and the volume, i.e., $\rho(T,V)$  \cite{Tamayo}, we have the following relation:
\begin{eqnarray} \label{ecuacionparaT}
\frac{d \rho}{d a}=\frac{\partial \rho}{\partial T}\frac{d T}{d a}+\frac{3n}{a}\frac{\partial \rho}{\partial n}.
\end{eqnarray}
Using Eqs. \eqref{fluidoperfecto}, \eqref{ecuacionparaT}, and the EoS, it can be shown (as in \cite{Tamayo}) that the temperature from \eqref{temperatura} is directly proportional to the internal energy, and is given by
\begin{equation}\label{Tperfecto}
    \frac{T}{T_{0}}=\frac{\rho}{\rho_{0}}a^{3}=\frac{U}{U_{0}}.
\end{equation}
Additional to this, from Eq. \eqref{gibbspaso1}, together with Eqs. \eqref{npunto}, \eqref{fluidoperfecto}, and the EoS, we have $dS=0$ or, consequently, $dS/dt=0$, which imply that there is no entropy production in the cosmic expansion, i.e. the fluid is adiabatic.

For a viscous fluid an average 4-velocity is chosen in which there is no particle flux \cite{Eckart,librocaro}; so, in this frame, the particle 4-current is taking again as $n^{\alpha}_{;\alpha}=0$ and the equation \eqref{npunto} is still valid. On the other hand, from the  Eckart's theory, we have the following conservation equation, according to Eqs.\eqref{ConsEq1}, and \eqref{Peff}
\begin{equation}\label{ConsEq}
 \dot{\rho}+3H(\rho+p+\Pi)=0,
\end{equation}
which together with the Eq. \eqref{npunto} and the EoS \eqref{gibbspaso1}, give us the follow expression for the entropy production \cite{Tamayo,entropiaprofe}

\begin{equation}\label{entropyproduction} 
    \dfrac{d S}{d \tau}=\frac{3E^{2}\Omega_{\xi_{0}}\rho}{nT},
\end{equation}
written in a dimensionless form. Therefore, the entropy production in the viscous expanding universe is, in principle, always positive and we recovered the behavior of a perfect fluid when $\Omega_{\xi_{0}}=0$. As we will see later, this positiveness requires some constraints under the free parameters of the solution, specifically, in the expression for the temperature of the dissipative fluid.

\section{Study of the exact solution}\label{exactlate}

In this section we study the exact solution \eqref{Tm1} under the condition \eqref{condicionBigRip} in terms of the fulfillment at the same time of the near equilibrium condition and the positiveness of the entropy production. For that end, we focus our analysis in two defined late-time epochs of validity for the solution: (i) the actual time $\tau=0$ for which $E=1$, and (ii) the very late-times $\tau\to\infty$ for which $E=\sqrt{\Omega_{\Lambda_{0}}}$. We will extend the analysis for $\tau>0$ and for $\tau<0$. 

It is important to mention that the asymptotic behavior of the solution \eqref{Tm1}, given by the de Sitter solution \eqref{HdeSitter} when the condition \eqref{condicionBigRip} holds, leads to a universe dominated only by the CC in which the dissipative WDM is diluted due to the universe expansion, as can be seen by evaluating the Eq. \eqref{HdeSitter} in the Friedmann equation \eqref{tt}

\begin{equation}\label{tt1}
H^{2}_{dS}=\frac{\rho}{3}+\frac{\Lambda}{3},
\end{equation}
which leads to $\rho=0$. Therefore, in this asymptotic behavior we do not have a dissipative fluid to study the near equilibrium condition and the entropy production. Nevertheless, we can study these two conditions as an asymptotic regime of the solution while $\rho\to 0$.

\subsection{Near equilibrium condition of the exact solution}\label{seccion4A}
Note that the near equilibrium condition \eqref{lhm1}, considering that $1<\gamma\leq 2$, can be rewritten as
\begin{equation}\label{gammaWDM}
  E(\tau)\ll\frac{\gamma-1}{\Omega_{\xi_{0}}}=\frac{\gamma}{\Omega_{\xi_{0}}}-\frac{1}{\Omega_{\xi_{0}}},
\end{equation}
expression that tells us that as a long as $E=H/H_{0}$ be much smaller than $(\gamma-1)/\Omega_{\xi_{0}}$, the solution must be near to the thermodynamic equilibrium. This open the possibility that the solution is able to fulfill this condition considering that $E(t)$, given by Eq. \eqref{Tm1}, decrease asymptotically to the de Sitter solution \eqref{HdeSitter} as $\tau\rightarrow\infty$ under the condition \eqref{condicionBigRip}, as we can be seen in Fig. \ref{figuraem1}. Furthermore, at the actual time $\tau=0$, the condition \eqref{gammaWDM} leads to
\begin{equation} \label{actualnearcondition}
    \Omega_{\xi_{0}}\ll\gamma-1,
\end{equation}
and for the very late times $\tau\to\infty$ leads to
\begin{equation} \label{latenearcondition}
    \sqrt{\Omega_{\Lambda_{0}}}\ll\frac{\gamma-1}{\Omega_{\xi_{0}}}=\frac{\gamma}{\Omega_{\xi_{0}}}-\frac{1}{\Omega_{\xi_{0}}}.
\end{equation}

The fulfillment of the condition \eqref{actualnearcondition} implies the fulfillment of the condition \eqref{latenearcondition}, because $0<\Omega_{\Lambda_{0}}\leq 1$ and from the condition \eqref{actualnearcondition} we get $1\ll(\gamma-1)/\Omega_{\xi_{0}}$. Also, note that the fulfilment of the condition \eqref{actualnearcondition} implies the fulfilment of the condition \eqref{condicionBigRip}. Summarizing, the fulfillment of the condition \eqref{actualnearcondition} implies the fulfillment of the near equilibrium condition from $0\leq\tau<\infty$ and the condition \eqref{condicionBigRip} for which the solutions \eqref{Tm1} behave as the de Sitter solution at very late times. It is important to note that, the fulfilment of the near equilibrium condition depends only on the values of $\gamma$ and $\Omega_{\xi_{0}}$ and not in the values of $\Omega_{\Lambda_{0}}$, with the characteristic that for a value of $\gamma$ more closer to $1$ (CDM) a smaller value of $\Omega_{\xi_{0}}$ must be considered. Even more, for $1<\gamma\leq 2$ we can see that $\Omega_{\xi_{0}}<1$. On the other hand, the condition \eqref{actualnearcondition} is independent of the behavior of the solution because the election $E(\tau=0)=1$ is arbitrary, but, this not implies that the condition \eqref{latenearcondition} be always fulfilled because this condition depends on the behavior of the solution. In this sense note that if we do not satisfy the condition \eqref{condicionBigRip} in Eq. \eqref{lhm1}, then $E$ diverges and the solution will be far from near equilibrium in a finite time in the Big-Rip scenario.

\begin{figure}[ht]
\includegraphics[width=0.48\textwidth]{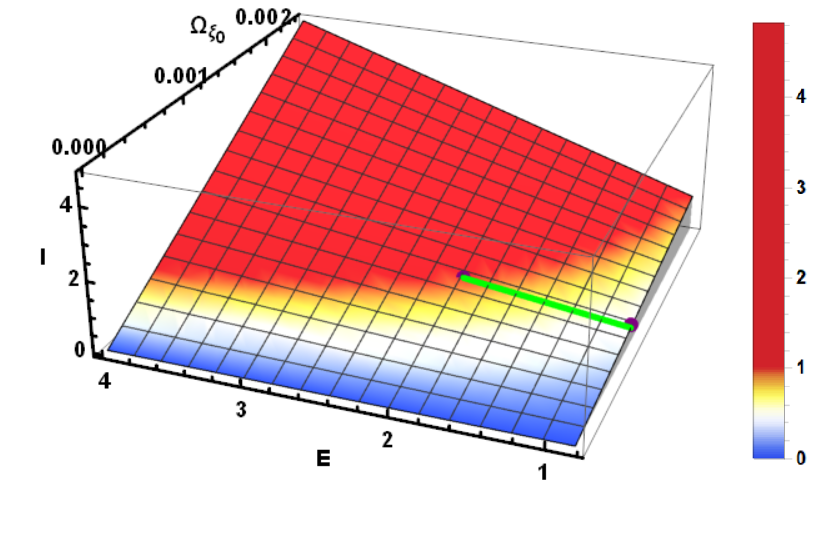}
\caption{Behavior of $l$, given by Eq. \eqref{lhm1}, for $0\leq\Omega_{\xi_{0}}\leq0.02$ and $\sqrt{\Omega_{\Lambda_{0}}}\leq E\leq 4$. We also consider the fixed values of $\Omega_{\Lambda_{0}}=0.69$ and $\gamma=1.002$. The green line represent the near equilibrium condition when $\Omega_{\xi_{0}}=0.001$ and the red zone represent the values for which we are far from the near equilibrium condition ($l>1$).}
\label{historianear}
\end{figure}

For $\tau<0$ it is still possible to fulfill the near equilibrium condition \eqref{gammaWDM}, as we mentioned above, while $E$ be much smaller than $(\gamma-1)/\Omega_{\xi_{0}}$. In Fig. \ref{historianear} we depict the near equilibrium condition \eqref{lhm1} as a function of $\Omega_{\xi_{0}}$ and $E$ for the fixed values of $\Omega_{\Lambda_{0}}=0.69$ and $\gamma=1.002$. The red zone represent the values for which we are far from the near equilibrium and the green line represent the near equilibrium condition when $\Omega_{\xi_{0}}=0.001$. Note that, for these last values we are far from the equilibrium when $\Omega_{\xi_{0}}\geq 0.002$ ($l\geq 1$) for all $E$, and for $\Omega_{\xi_{0}}=0.001$ we are far from the near equilibrium when $E\geq 2$, i. e., we are far from the near equilibrium condition for a time given by $\tau=-0.6001$ which is roughly equivalent to $8.63389$ Gyrs backward in time (0.630211 times the life of the $\Lambda$CDM universe), according to the Eq. \eqref{Tm1}. 

From Fig. \ref{historianear}  we can see that when $E$ grows, we can make $\Omega_{\xi_{0}}$ more closer to zero in order to fulfill the near equilibrium condition. For the solution \eqref{Tm1} under the condition \eqref{condicionBigRip} this means that for $\tau<\infty$, for which $E$ grows, we can stay in the near equilibrium while $\Omega_{\xi_{0}}$ be small enough to satisfy the condition \eqref{gammaWDM}. This is due to the election of the dissipation of the form $\xi=\xi_{0}\rho$, because the dissipative pressure $\Pi=-3H\xi$ behaves as $\rho^{3/2}$ and the equilibrium pressure behaves as $\rho$ and, therefore, considering that in this case $\rho$ grows to the past, then the dissipative pressure grows more quickly than the equilibrium pressure and $\xi_{0}$ acts as a modulator of this growth.

\subsection{Entropy production of the exact solution}\label{seccion4C}

In order to obtain the entropy production of the dissipative fluid from Eq. \eqref{entropyproduction}, we need to find their temperature from Eq. \eqref{temperatura}. Rewriting the conservation Eq. \eqref{ConsEq} in the form
\begin{equation}\label{pararhoa}
    \frac{d\rho}{da}=-\frac{3\rho}{a}\left(\gamma-3H\xi_{0}\right),
\end{equation}
we can rewrite Eq. \eqref{ecuacionparaT} as
\begin{equation}\label{paso1}
    \rho\left(\gamma-3H\xi_{0}\right)=-\frac{a}{3}\frac{\partial \rho}{\partial T}\frac{dT}{da}-n\frac{\partial \rho}{\partial n}.
\end{equation}
Then, from Eq. \eqref{temperatura}, and using Eq. \eqref{Peff}, we have
\begin{equation}\label{paso2}
    n\frac{\partial T}{\partial n}+\rho \left(\gamma-3H\xi_{0}\right)\frac{\partial T}{\partial \rho}=T\left[\left(\gamma-1\right)-3H\xi_{0}-3\xi_{0}\rho\frac{\partial H}{\partial \rho}\right],
\end{equation}
an expression that, with together Eq. \eqref{paso1} leads to
\begin{equation}\label{paso8}
\frac{dT}{T}=-3\frac{da}{a}\left[\left(\gamma-3H\xi_{0}\right)-1-3\xi_{0}\rho\frac{\partial H}{\partial \rho}\right].   
\end{equation}
Thus, using Eqs. \eqref{tt} and \eqref{pararhoa} we get, from Eq. \eqref{paso8}, the following expression:
 \begin{equation}
\frac{d T}{T}=\frac{d\rho}{\rho}\left[1-\frac{\left(\frac{2}{3}\sqrt{3\left(\rho+\Lambda\right)}+\xi_{0}\rho\right)}{\frac{2}{3}\sqrt{3\left(\rho+\Lambda\right)}\left(\gamma-\sqrt{3\left(\rho+\Lambda\right)}\xi_{0}\right)}\right],
 \end{equation}
which has the following solution in our dimensionless notation:
\begin{eqnarray} 
&\ln{\left(\frac{T}{T_{0}}\right)}=\ln\left(\frac{\rho}{\rho_{0}}\right)\nonumber\\
&+\frac{2\Omega_{\xi_{0}}\sqrt{\Omega_{\Lambda}}\left[\arctanh{\left(\frac{\sqrt{\Omega_{\Lambda_{0}}}}{E}\right)}-\arctanh{\left(\sqrt{\Omega_{\Lambda_{0}}}\right)}\right]}{\left(\gamma^{2}-\Omega_{\Lambda_{0}}\Omega^{2}_{\xi_{0}}\right)}\nonumber\\ \label{paso3}
&\frac{-\gamma\ln\left({\frac{\rho}{\rho_{0}}}\right)+\left[\gamma(2+\gamma)-\Omega_{\Lambda_{0}}\Omega^{2}_{\xi_{0}}\right]\ln{\left(\frac{\gamma-E\Omega_{\xi_{0}}}{\gamma-\Omega_{\xi_{0}}}\right)}}{\left(\gamma^{2}-\Omega_{\Lambda_{0}}\Omega^{2}_{\xi_{0}}\right)}.
\end{eqnarray}

On the other hand, integrating Eq. \eqref{pararhoa} with the help of Eq. \eqref{tt}, we obtain in our dimensionless notation the expression
\begin{eqnarray}
\ln{a^{3}}&=&\frac{2\Omega_{\xi_{0}}\sqrt{\Omega_{\Lambda}}\left[\arctanh{\left(\frac{\sqrt{\Omega_{\Lambda_{0}}}}{E}\right)}-\arctanh{\left(\sqrt{\Omega_{\Lambda_{0}}}\right)}\right]}{\left(\gamma^{2}-\Omega_{\Lambda_{0}}\Omega^{2}_{\xi_{0}}\right)}\nonumber\\ \label{paso4}
&+&\frac{-\gamma\ln\left({\frac{\rho}{\rho_{0}}}\right)+2\gamma\ln{\left(\frac{\gamma-E\Omega_{\xi_{0}}}{\gamma-\Omega_{\xi_{0}}}\right)}}{\left(\gamma^{2}-\Omega_{\Lambda_{0}}\Omega^{2}_{\xi_{0}}\right)},
\end{eqnarray}
from which we can express the Eq. \eqref{paso3} in terms of the scale factor as follows:
\begin{equation}
    \ln{\left(\frac{T}{T_{0}}\right)}=\ln\left(\frac{\rho}{\rho_{0}}\right)+\ln\left(\frac{\gamma-E\Omega_{\xi_{0}}}{\gamma-\Omega_{\xi_{0}}}\right)+\ln{a^{3}}.
\end{equation}
Hence, the temperature of the dissipative fluid as a function of the scale factor is given by
\begin{equation}\label{temperaturalate}
    T=T_{0}\left(\frac{\rho}{\rho_{0}}\right)\left(\frac{\gamma-E\Omega_{\xi_{0}}}{\gamma-\Omega_{\xi_{0}}}\right){a^{3}},
\end{equation}
which reduced to the expression for the temperature of a perfect fluid, given by Eq. \eqref{Tperfecto}, when $\Omega_{\xi_{0}}=0$.

Note that it is possible to obtain an expression for the temperature of the dissipative fluid that does not depend on $\rho$, by combining the Eqs. \eqref{paso4} and \eqref{temperaturalate}, which leads to
\begin{equation}\label{Treal}
\begin{split}
     T&=T_{0}a^{\frac{3}{\gamma}\left(\gamma-\gamma^2+{\Omega^2_{\xi_{0}}\Omega_{\Lambda_{0}}}\right)}\left(\frac{\gamma-E\Omega_{\xi_{0}}}{\gamma-\Omega_{\xi_{0}}}\right)^3\times\\
     & \left[\left(\frac{E+\sqrt{\Omega_{\Lambda_{0}}}}{E-\sqrt{\Omega_{\Lambda_{0}}}}\right)\left(\frac{1-\sqrt{\Omega_{\Lambda_{0}}}}{1+\sqrt{\Omega_{\Lambda_{0}}}}\right)\right]^{\left(\frac{\Omega_{\xi_{0}}\sqrt{\Omega_{\Lambda_{0}}}}{\gamma}\right)},
\end{split}
\end{equation}
and from which we can see that the temperature is always positive in two cases: (i) when $\gamma-E\Omega_{\xi_{0}}>0$ and $\gamma-\Omega_{\xi_{0}}>0$ or (ii) when $\gamma-E\Omega_{\xi_{0}}<0$ and $\gamma-\Omega_{\xi_{0}}<0$. Therefore, if we fulfill the near equilibrium condition \eqref{gammaWDM}, then we obtain a positive expression for the temperature, since the case (i), and from the same condition given by Eq. \eqref{gammaWDM}, we can also fulfill the condition \eqref{condicionBigRip} from which the solution \eqref{Tm1} tends asymptotically to the future at the usual de Sitter solution. Note that the case (ii) implies that the fluid is far from the near equilibrium and the solution \eqref{Tm1} has a Big-Rip singularity. On the other hand, considering that the solution \eqref{Tm1} is a decreasing function with time when the condition \eqref{condicionBigRip} holds, then the cubic term in the Eq. \eqref{Treal} is also a decreasing function; thus, considering that $E(\tau\to\infty)\to\sqrt{\Omega_{\Lambda_{0}}}$, a decreasing temperature with time requires that
\begin{equation}
    \frac{a^{\frac{3}{\gamma}\left(\gamma-\gamma^{2}+\Omega_{\xi_{0}}^{2}\Omega_{\Lambda_{0}}\right)}}{\left(E-\sqrt{\Omega_{\Lambda}}\right)^{\left(\frac{\Omega_{\xi_{0}}\sqrt{\Omega_{\Lambda_{0}}}}{\gamma}\right)}}\to 0,
\end{equation}
which is only possible, considering that $a(\tau\to\infty)\to\infty$, when the exponent of the power law for the scale factor is negative, i. e., if
\begin{equation}
    \Omega_{\Lambda_{0}}<\frac{\gamma}{\Omega_{\xi_{0}}}\frac{(\gamma-1)}{\Omega_{\xi_{0}}},
\end{equation}
which is always true because $0<\Omega_{\Lambda_{0}}\leq 1$ and the fulfilment of the condition \eqref{condicionBigRip} implies that $1<\gamma/\Omega_{\xi_{0}}$, as well as the fulfillment of the condition \eqref{actualnearcondition} implies that $1\ll (\gamma-1)/\Omega_{\xi_{0}}$. It is important to note that for a CDM ($\gamma=1$) the exponent of the power law for the scale factor is always positive and the temperature is an increasing function with time, which represents a contradictory behavior for an expanding universe.

\begin{figure}[H]
\centering
\includegraphics[width=0.48\textwidth]{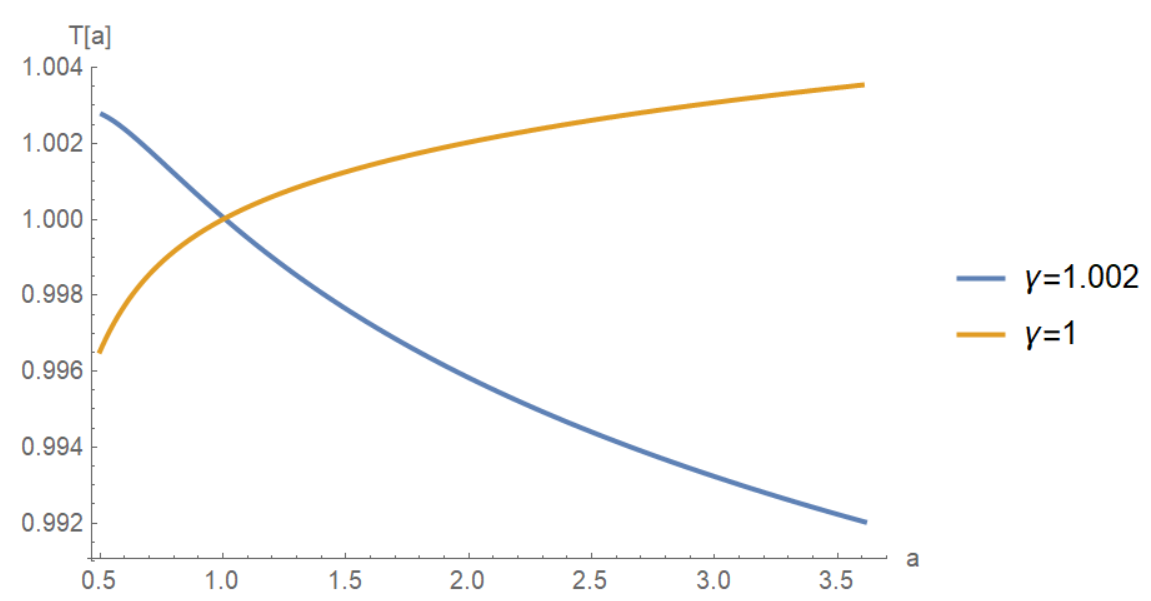}
\caption{Numerical behavior of $T$, given by Eq. \eqref{temperaturalate}, for $0.5\leq a\leq 3.5$. We also consider the fixed values of $T_{0}=1$, $\Omega_{\xi_{0}}=0.001$, and $\Omega_{\Lambda_{0}}=0.69$.}
\label{figuraTlate}
\end{figure}

In the Fig. \ref{figuraTlate} we present the numerical behavior of temperature $T$ of the dissipative fluid, given by the Eq. \eqref{temperaturalate}, as a function of the scale factor $a$. We rewrite $E$ as a function of the energy density $\rho$ from Eq. \eqref{tt} and we use the expression for $\rho$ given by the Eq. \eqref{paso4}. For the free parameters we use the values of $T_{0}=1$, $\Omega_{\xi_{0}}=0.001$, $\Omega_{\Lambda_{0}}=0.69$, and $\gamma=1.002$. We also present the behavior of the temperature when $\gamma=1$. It is important to mention that the difference between the initial value of the temperature for the WDM case and his final value is $0.0102895$ for a scale factor that is $3.6$ times more bigger than the actual size of the Universe, this is the result to be close to the near equilibrium condition, that makes that temperature decrease very slowly to zero.

With the temperature of the dissipative fluid given by Eq. \eqref{temperaturalate}, we can calculate the entropy from Eq. \eqref{entropyproduction}, using also that from Eq. \eqref{npunto}, we have $n=n_{0}a^{-3}$. Then, Eq. \eqref{entropyproduction} in our dimensionless notation, takes the following form:
\begin{eqnarray}\label{entropiapaso1}
\frac{dS}{d \tau}=\frac{3E^{2}\Omega_{\xi_{0}}\rho}{nT}=\frac{3E^{2}\rho_{0}\Omega_{\xi_{0}}\left(\gamma-\Omega_{\xi_{0}}\right)}{n_{0}T_{0}\left(\gamma-E\Omega_{\xi_{0}}\right)}.
\end{eqnarray}
The positiveness of the entropy production depends on the same cases as the positiveness of the temperature of the dissipative fluid. Therefore, the fulfillment of the near equilibrium condition, implies the positiveness of the entropy production. Note that for the asymptotic de Sitter solution the entropy production goes to a constant but, in this case, $\rho\rightarrow0$ (which implies a null $\Omega_{\xi_{0}}$) and then we have a null entropy production. On the other hand, as well as the temperature of the dissipative fluid, the entropy production goes to infinity when we do not satisfy the condition \eqref{condicionBigRip} in a finite time in the Big-Rip singularity. 

\begin{figure}[H]
\centering
\includegraphics[width=0.48\textwidth]{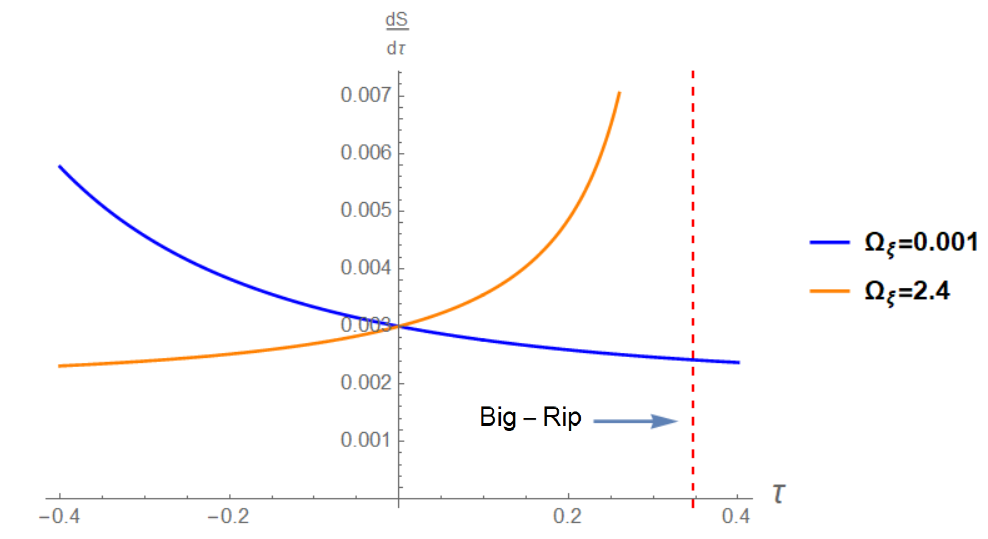}
\caption{Numerical behavior of $dS/d\tau$, given by Eq. \eqref{entropiapaso1}, for $-0.4\leq\tau\leq 0.4$. We also consider the fixed values of $\Omega_{\Lambda_{0}}=0.69$ and $\gamma=1.002$. The red dashed line represent the time $\tau_{s}$, given by Eq. \eqref{TiempoBR}, in which the Big-Rip singularity occurs.}
\label{dsdt}
\end{figure}

In Fig. \ref{dsdt} we show the numerical behavior of the entropy production, given by the Eq. \eqref{entropiapaso1}, as a function of time $\tau$ for $\Omega_{\xi_{0}}=0.001$, $\Omega_{\Lambda_{0}}=0.69$, and $\gamma=1.002$. We also show the behavior of the entropy production when $\Omega_{\xi_{0}}=2.4$. For this last one, the near equilibrium condition is not satisfied, since this value enter in to contradiction with Eq. \eqref{condicionBigRip}, and the entropy production diverge in a finite time given by  \cite{primerarticulo}
\begin{equation}\label{TiempoBR}
\tau_{s}=\frac{2\Omega_{\xi_{0}}\log \left[ \left(\frac{1-\sqrt{\Omega_{\Lambda_{0}}}}{1+\sqrt{\Omega_{\Lambda_{0}}}}\right)^{\frac{\gamma}{2\Omega_{\xi_{0}}\sqrt{\Omega_{\Lambda_{0}}}}}\left(1-\Omega_{\Lambda_{0}}\right)^{\frac{1}{2}}\left(\frac{-\Omega_{\xi_{0}}}{\gamma-\Omega_{\xi_{0}}}\right)\right]}{3\left(\gamma^{2}-\Omega^{2}_{\xi_{0}}\Omega_{\Lambda_{0}}\right)},   
\end{equation}
which according to our parameters, this is equal to $\tau_{s}=0.346571$, which is roughly equivalent to $4.98626$Gyrs from the present time.
In summary, the condition given by Eq. \eqref{actualnearcondition} 
together with the condition  $\Omega_{\xi_{0}}<1$, for the present time, describe a viscous WDM model that is compatible with the near equilibrium condition, and presents a proper physical behavior of the temperature (a decreasing function with the scale factor), and entropy production (without future Big-Rip singularity). Therefore, in this sense, all this previous thermodynamics analysis will help us to define the best prior definition for our cosmological parameters $\gamma$ and $\Omega_{\xi_{0}}$, in order to constraints with the cosmological data. Accordingly, our model has two more free parameters than the standard $\Lambda$CDM model, which appear from the possibility of a more general model of the DM component, that takes into account a warm nature with a non-perfect fluid description.

\section{Cosmological constraints}\label{constraints}
In this section we shall constraint the free parameters of the viscous $\Lambda$WDM model with the SNe Ia data coming from the Pantheon sample \cite{Scolnic_Complete_2018}, which consists in $1048$ data points in the redshift range $0.01\leq z\leq 2.3$; and the OHD compiled by Magaña \textit{et al.} \cite{Magana_Cardassian_2018}, which consists in $51$ data points in the redshift range $0.07\leq z\leq 2.36$. To do so, we compute the best-fit parameters and their respective confidence regions with the affine-invariant Markov Chain Monte Carlo (MCMC) method \cite{Goodman_Ensemble_2010}, implemented in the pure-Python code \textit{emcee} \cite{Foreman_emcee_2013}, by setting $30$ chains or ``walkers''. As a convergence test, we compute every $50$ steps the autocorrelation time of the chains $\tau_{corr}$, provided by the \textit{emcee} module. If the current step is larger than $50\tau_{corr}$, and if the values of $\tau_{corr}$ changed by less than $1\%$, then we will consider that the chains are converged and the code is stopped. The first $5\tau_{corr}$ steps are discarded as ``burn-in'' steps. This convergence test is complemented with the mean acceptance fraction, which should be between $0.2$ and $0.5$ \cite{Foreman_emcee_2013}, and can be modified by the stretch move provided by the \textit{emcee} module.

Since we are implementing a Bayesian statistical analysis, we need to construct the Gaussian likelihood
\begin{equation}\label{likelihood}
    \mathcal{L}=\mathcal{N}\exp{\left(-\frac{\chi_{I}^{2}}{2}\right)}.
\end{equation}
Here, $\mathcal{N}$ is a normalization constant, which does not influence in the MCMC analysis, and $\chi_{I}^{2}$ is the merit function, where $I$ stands for each data set considered, namely, SNe Ia, OHD, and their joint analysis in which $\chi_{joint}^{2}=\chi_{SNe}^{2}+\chi_{OHD}^{2}$.

The merit function for the OHD data is constructed as
\begin{equation}\label{meritOHD}
    \chi_{OHD}^{2}=\sum_{i=1}^{51}{\left[\frac{H_{i}-H_{th}(z_{i},\theta)}{\sigma_{H,i}}\right]^{2}},
\end{equation}
where $H_{i}$ is the observational Hubble parameter at redshift $z_{i}$ with an associated error $\sigma_{H,i}$, all of them provided by the OHD sample, $H_{th}$ is the theoretical Hubble parameter at the same redshift, and $\theta$ encompasses the free parameters of the model under study. It is important to mention that in our MCMC analysis we consider the value of the Hubble parameter at the current time, $H_{0}$, as  a free parameter, which is written as $H_{0}=100\frac{km/s}{Mpc}h$, with $h$ dimensionless and for which we consider the Gaussian prior $G(0.7403,0.0142)$, according to the value of $H_{0}$ obtained by A. G. Riess \textit{et al.} \cite{Riess:2019cxk}.

On the other hand, the merit function for the SNe Ia data is constructed as
\begin{equation}\label{meritSNeIa}
    \chi_{SNe}^{2}=\sum_{i=1}^{1048}{\left[\frac{\mu_{i}-\mu_{th}(z_{i},\theta)}{\sigma_{\mu,i}}\right]^{2}},
\end{equation}
where $\mu_{i}$ is the observational distance modulus of each SNe Ia at redshift $z_{i}$ with an associated error $\sigma_{\mu,i}$, $\mu_{th}$ is the theoretical distance modulus for each SNe Ia at the same redshift, and $\theta$ encompasses the free parameters of the model under study. Following this line, the theoretical distance modulus can be obtained, for a flat FLRW space-time, from the expression
\begin{equation}\label{theoreticaldistance}
    \mu_{th}(z_{i},\theta)=5\log_{10}{\left[\frac{d_{L}(z_{i},\theta)}{Mpc}\right]}+\Bar{\mu},
\end{equation}
where $\Bar{\mu}=5\left[\log_{10}{\left(c\right)}+5\right]$, $c$ is the speed of light given in units of $km/s$, and $d_{L}$ is the luminosity distance given by
\begin{equation}\label{luminosity}
    d_{L}(z_{i},\theta)=(1+z_{i})\int_{0}^{z_{i}}{\frac{dz'}{H(z',\theta)}}.
\end{equation}

In the Pantheon sample the distance estimator is obtained using a modified version of the Tripp's formula \cite{Tripp_Luminosity_1998}, with two nuisance parameters calibrated to zero with the BEAMS whit Bias Correction (BBC) method \cite{Kessler_Correcting_2017}. Hence, the observational distance modulus for each SNe Ia is given by
\begin{equation}\label{pantheondistance}
    \mu_{i}=m_{B,i}-\mathcal{M},
\end{equation}
where $m_{B,i}$ is the corrected apparent B-band magnitude of a fiducial SNe Ia at redshift $z_{i}$, all of them provided by the pantheon sample \footnote{Available online in the GitHub repository \textit{https://github.com/dscolnic/Pantheon}. The corrected apparent B-band magnitude $m_{B,i}$ for each SNe Ia with their respective redshifts $z_{i}$ and errors $\sigma_{m_{B},i}$ are available in the document \textit{lcparam\_full\_long.txt}.}, and $\mathcal{M}$ is a nuisance parameter which must be jointly estimated with the free parameters $\theta$ of the model under study. Therefore, we can rewrite the merit function \eqref{meritSNeIa} in matrix notation (denoted by bold symbols) as
\begin{equation}\label{meritSNeIamatrix}
    \chi_{SNe}^{2}=\mathbf{M}(z,\theta,\mathcal{M})^{\dagger}\mathbf{C}^{-1}\mathbf{M}(z,\theta,\mathcal{M}),
\end{equation}
where $[\mathbf{M}(z,\theta,\mathcal{M})]_{i}=m_{B,i}-\mu_{th}(z_{i},\theta)-\mathcal{M}$, and $\mathbf{C}=\mathbf{D}_{stat}+\mathbf{C}_{sys}$ is the total uncertainties covariance matrix, being $\mathbf{D}_{stat}=diag(\sigma_{m_{B},i}^{2})$ the statistical uncertainties of $m_{B}$ and $\mathbf{C}_{sys}$ the systematic uncertainties in the BBC approach \footnote{Available online in the GitHub repository \textit{https://github.com/dscolnic/Pantheon} in the document \textit{sys\_full\_long.txt}.}.

Finally, to marginalize over the nuisance parameters $\Bar{\mu}$ and $\mathcal{M}$, we define $\Bar{\mathcal{M}}=\Bar{\mu}+\mathcal{M}$, and the merit function \eqref{meritSNeIamatrix} is expanded as \cite{Lazkos_Exploring_2005}
\begin{equation}\label{meritSNeIaexpanded}
    \chi_{SNe}^{2}=A(z,\theta)-2B(z,\theta)\Bar{\mathcal{M}}+C\Bar{\mathcal{M}}^{2},
\end{equation}
where
\begin{equation}\label{defofA}
    A(z,\theta)=\mathbf{M}(z,\theta,\Bar{\mathcal{M}}=0)^{\dagger}\mathbf{C}^{-1}\mathbf{M}(z,\theta,\Bar{\mathcal{M}}=0),
\end{equation}
\begin{equation}\label{defofB}
    B(z,\theta)=\mathbf{M}(z,\theta,\Bar{\mathcal{M}}=0)^{\dagger}\mathbf{C}^{-1}\mathbf{1},
\end{equation}
\begin{equation}
    C=\mathbf{1}\mathbf{C}^{-1}\mathbf{1}.
\end{equation}
Therefore, by minimizing the expanded merit function \eqref{meritSNeIaexpanded} with respect to $\Bar{\mathcal{M}}$, it is obtained $\Bar{\mathcal{M}}=B(z,\theta)/C$, and the expanded merit function reduced to
\begin{equation}\label{meritSNeIaminimize}
    \chi_{SNe}^{2}=A(z,\theta)-\frac{B(z,\theta)^{2}}{C},
\end{equation}
which depends only on the free parameters of the model under study. 

It is important to mention that the expanded and minimized merit function \eqref{meritSNeIaminimize} provides the same information as the merit function \eqref{meritSNeIamatrix} since the best-fit parameters minimize the merit function, and their corresponding value can be used as an indicator of the goodness of the fit independently of the data set used: the smaller the value of $\chi_{min}^{2}$ is, the better is the fit. In this sense, in principle the value of $\chi_{min}^{2}$ obtained for the best fit parameters can be reduced by adding free parameters to the model under study, resulting in overfitting. Hence, we compute the Bayesian criterion information (BIC) \cite{Schwarz_Estimating_1978} to compare the goodness of the fit statistically. This criterion adds a penalization in the value of $\chi_{min}^{2}$ that depends on the total number of free parameters of the model, $\theta_{N}$, according to the expression
\begin{equation}\label{BIC}
    BIC=\theta_{N}\ln{(n)}+\chi_{min}^{2},
\end{equation}
where $n$ is the total number of data points in the corresponding data sample. So, when two different models are compared, the one most favored by the observations statistically corresponds to the one with the smallest value of BIC. In general, a difference of $2-6$ in BIC is evidence against the model with higher BIC,  a difference of $6-10$ is strong evidence, and a difference $>10$ is very strong evidence. 

Since in the merit function of the two data sets, the respective model is considered thought the Hubble parameter as a function of the redshift (see Eqs. \eqref{meritOHD} and \eqref{luminosity}), then we numerically integrate Eq. \eqref{Hpunto} with $m=1$, which can be rewritten, considering that $\dot{z}=-(1+z)H$, as
\begin{equation}\label{Hconstraint}
    \dfrac{dH}{dz}=\frac{1}{2(1+z)}\left[3\gamma H-3\xi_{0}\left(3H^{2}-\Lambda\right)-\frac{\Lambda\gamma}{H}\right],
\end{equation}
using as initial condition $H(z=0)=H_{0}=100\frac{km/s}{Mpc}h$, and taking into account that $\xi_{0}=\Omega_{\xi_{0}}/(3H_{0})$ and $\Lambda=3H_{0}^{2}(1-\Omega_{m0})$; this last one derived from Eq. \eqref{tt}, which leads to $\Omega_{m 0}+\Omega_{\Lambda 0}=1$. Even more, for a further comparison, we also constraint the free parameters of the $\Lambda$CDM model, whose respective Hubble parameter as a function of the redshift is given by
\begin{equation}\label{HLCDM}
    H(z)=100\frac{km/s}{Mpc}h\sqrt{\Omega_{m0}(1+z)^{3}+1-\Omega_{m0}}.
\end{equation}
Therefore, the free parameters of the viscous $\Lambda$WDM model are $\theta=\{h,\Omega_{m0},\Omega_{\xi_{0}},\gamma\}$, and the free parameters of the $\Lambda$CDM model are $\theta=\{h,\Omega_{m0}\}$. For the free parameters $\Omega_{m0}$, $\Omega_{\xi_{0}}$, and $\gamma$ we consider the following priors: $\Omega_{m0}\in F(0,1)$, $\gamma\in G(1.00,0.02)$, and $0<\Omega_{\xi_{0}}<\gamma-1$; where $F$ stands for flat prior, and the prior for $\Omega_{\xi_{0}}$ is derived from the constraint given by Eq. \eqref{actualnearcondition}.

In Table \ref{tab:emcee} we present the total steps, the mean acceptance fraction (MAF), and the autocorrelation time $\tau_{corr}$ of each free parameter, obtained when the convergence test is fulfilled during our MCMC analysis for both, the viscous $\Lambda$WDM and $\Lambda$CDM, models. The values of the MAF are obtained for a value of the stretch move of $a=7$ for the $\Lambda$CDM model, and a value of $a=3$ for the viscous $\Lambda$WDM model.

\begin{table}[]
    \centering
    \caption{Final values of the total number of steps, mean acceptance fraction (MAF), and autocorrelation time $\tau_{corr}$ for each model free parameters, obtained when the convergence test described in Section \ref{constraints} is fulfilled for a MCMC analysis with $30$ chains or ``walkers''. The values of the MAF are obtained for a value of the stretch move of $a=7$ for the $\Lambda$CDM model, and a value of $a=3$ for the viscous $\Lambda$WDM model.}
    \begin{tabular}{ccccccc}
    \hline\hline
         & & & \multicolumn{4}{c}{$\tau_{corr}$} \\
        \cline{4-7}
        Data & Total steps & MAF & $h$ & $\Omega_{m0}$ & $\Omega_{\xi_{0}}$ & $\gamma$ \\
         \hline
         \multicolumn{7}{c}{$\Lambda$CDM model} \\ \\
        SNe Ia & $1050$ & $0.370$ & $16.5$ & $17.5$ & $\cdots$ & $\cdots$ \\
        OHD & $1000$ & $0.367$ & $14.9$ & $17.1$ & $\cdots$ & $\cdots$ \\
        SNe Ia+OHD & $800$ & $0.364$ & $15.8$ & $15.4$ & $\cdots$ & $\cdots$ \\
        \hline
        \multicolumn{7}{c}{viscous $\Lambda$WDM model} \\ \\
        SNe Ia & $2700$ & $0.385$ & $45.8$ & $44.3$ & $49.6$ & $51.9$ \\
        OHD & $2450$ & $0.377$ & $39.6$ & $45.5$ & $48.2$ & $48.9$ \\
        SNe Ia+OHD & $2700$ & $0.379$ & $43.0$ & $45.9$ & $50.5$ & $53.3$ \\
        \hline\hline
    \end{tabular}
    \label{tab:emcee}
\end{table}

\section{Results and discussion}\label{results}
\begin{figure}
    \centering
    \includegraphics[scale = 0.35]{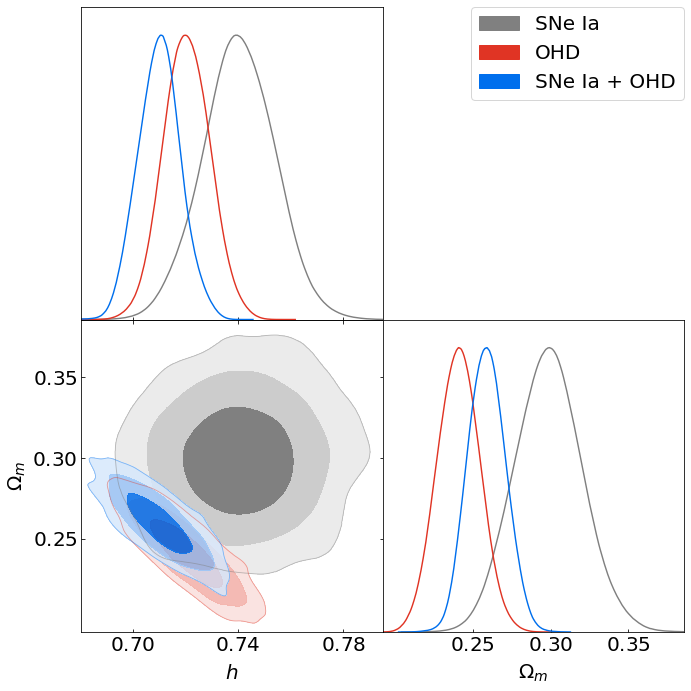}
    \caption{Joint and marginalized regions of the free parameters $h$ and $\Omega_{m0}$ for the $\Lambda$CDM model, obtained in the MCMC analysis described in the Section \ref{constraints}. The admissible regions presented in the joint regions correspond to $1\sigma(68.3\%)$, $2\sigma(95.5\%)$, and $3\sigma(99.7\%)$ of confidence level (CL), respectively. The best-fit values for each model free parameter are shown in Table \ref{tab:bestfits}.}
    \label{fig:TriangleLCDM}
\end{figure}

\begin{figure}
    \centering
    \includegraphics[scale = 0.35]{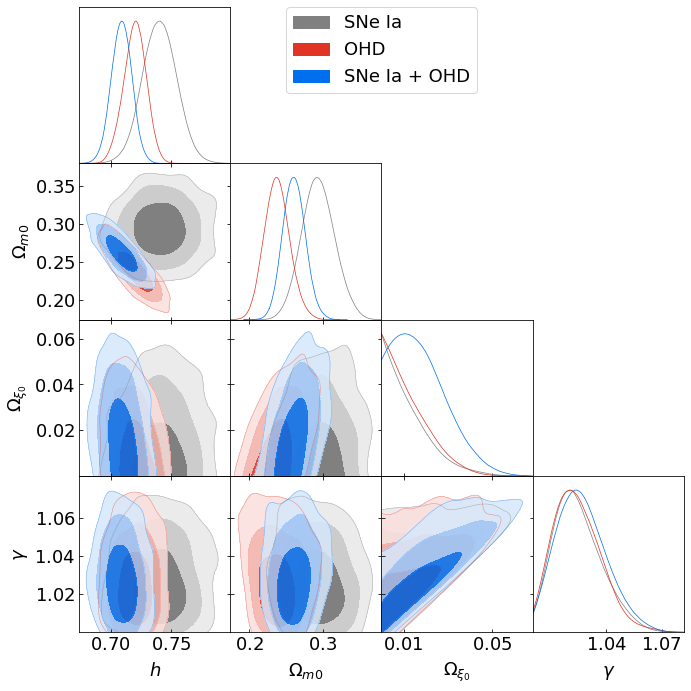}
    \caption{Joint and marginalized regions of the free parameters $h$, $\Omega_{m0}$, $\Omega_{\xi_{0}}$, and $\gamma$ for the viscous $\Lambda$WDM model, obtained in the MCMC analysis described in the Section \ref{constraints}. The admissible regions presented in the joint regions correspond to $1\sigma(68.3\%)$, $2\sigma(95.5\%)$, and $3\sigma(99.7\%)$ of confidence level (CL), respectively. The best-fit values for each model free parameter are shown in Table \ref{tab:bestfits}.}
    \label{fig:TriangleLWDM}
\end{figure}

The best-fit values for the $\Lambda$CDM and viscous $\Lambda$WDM models, obtained for the SNe Ia data, OHD, and in their joint analysis, as well as their corresponding goodness of fit criteria, are presented in Table \ref{tab:bestfits}. The uncertainties presented correspond to $1\sigma(68.3\%)$, $2\sigma(95.5\%)$, and $3\sigma(99.7\%)$ of confidence level (CL). In Figures \ref{fig:TriangleLCDM} and \ref{fig:TriangleLWDM} we depict the joint and marginalized credible regions of the free parameters space of the $\Lambda$CDM and viscous $\Lambda$WDM model, respectively. The admissible regions presented in the joint regions correspond to $1\sigma$, $2\sigma$, and $3\sigma$ CL.

\begin{table*}[]
    \centering
    \caption{Best-fit values and goodness of fit criteria for the viscous $\Lambda$WDM model with free parameters $h$, $\Omega_{m0}$, $\Omega_{\xi_{0}}$, and $\gamma$; and for the $\Lambda$CDM model with free parameters $h$ and $\Omega_{m0}$, obtained in the MCMC analysis described in the Section \ref{constraints} for the SNe Ia data, OHD, and in their joint analysis. The uncertainties correspond to $1\sigma(68.3\%)$, $2\sigma(95.5\%)$, and $3\sigma(99.7\%)$ of confidence level (CL), respectively. The best-fits values for the $\Lambda$CDM model are used for the sake of comparison with the viscous $\Lambda$WDM model.}
    \begin{tabular}{ccccccc}
        \hline\hline
         & \multicolumn{4}{c}{Best-fit values} & \multicolumn{2}{c}{Goodness of fit} \\
         \cline{2-7}
        Data & $h$ & $\Omega_{m0}$ & $\Omega_{\xi_{0}}(\times 10^{-2})$ & $\gamma$ & $\chi_{min}^{2}$ & $BIC$  \\
        \hline
        \multicolumn{7}{c}{$\Lambda$CDM model} \\ \\
        SNe Ia & $0.740_{-0.013\;-0.028\;-0.040}^{+0.014\;+0.028\;+0.043}$ & $0.299_{-0.022\;-0.042\;-0.058}^{+0.022\;+0.045\;+0.064}$ & $\cdots$ & $\cdots$ & $1026.9$ & $1040.8$ \\ \\
        OHD & $0.720_{-0.009\;-0.017\;-0.025}^{+0.009\;+0.018\;+0.026}$ & $0.241_{-0.014\;-0.027\;-0.037}^{+0.014\;+0.027\;+0.038}$ & $\cdots$ & $\cdots$ & $28.6$ & $36.5$ \\ \\
        SNe Ia+OHD & $0.710_{-0.008\;-0.016\;-0.022}^{+0.008\;+0.016\;+0.023}$ & $0.259_{-0.012\;-0.022\;-0.036}^{+0.012\;+0.024\;+0.034}$ & $\cdots$ & $\cdots$ & $1058.3$ & $1072.3$ \\
        \hline 
        \multicolumn{7}{c}{Viscous $\Lambda$WDM model} \\ \\
        SNe Ia & $0.741_{-0.014\;-0.028\;-0.044}^{+0.014\;+0.029\;+0.044}$ & $0.293_{-0.022\;-0.043\;-0.064}^{+0.023\;+0.048\;+0.066}$ & $0.980_{-0.716\;-0.943\;-0.979}^{+1.232\;+2.954\;+4.318}$ & $1.023_{-0.011\;-0.019\;-0.021}^{+0.015\;+0.031\;+0.045}$ & $1026.9$ & $1054.7$ \\ \\
        OHD & $0.721_{-0.010\;-0.020\;-0.029}^{+0.009\;+0.018\;+0.026}$ & $0.237_{-0.017\;-0.033\;-0.045}^{+0.017\;+0.037\;+0.053}$ & $1.026_{-0.738\;-0.988\;-1.019}^{+1.250\;+2.641\;+3.763}$ & $1.023_{-0.012\;-0.019\;-0.022}^{+0.014\;+0.030\;+0.048}$ & $28.5$ & $44.2$ \\ \\
        SNe Ia+OHD & $0.709_{-0.009\;-0.017\;-0.027}^{+0.009\;+0.017\;+0.026}$ & $0.261_{-0.015\;-0.030\;-0.041}^{+0.015\;+0.030\;+0.047}$ & $1.633_{-1.059\;-1.544\;-1.627}^{+1.381\;+2.871\;+4.186}$ & $1.026_{-0.013\;-0.021\;-0.024}^{+0.015\;+0.031\;+0.045}$ & $1056.9$ & $1084.9$ \\
        \hline\hline
    \end{tabular}
    \label{tab:bestfits}
\end{table*}

From the best-fit parameters presented in Table \ref{tab:bestfits} it is possible to see that there is no remarkable differences between the best-fit values for $h$ and $\Omega_{m0}$ obtained for the $\Lambda$CDM model and the viscous $\Lambda$WDM model. This is an expected behavior due to the similarity between the two models at late-times, as well to the past, considering the best-fit values obtained for $\Omega_{\xi_{0}}$. From the point of view of the goodness of fit criteria, we can conclude that the two models are able to describe the SNe Ia, OHD, and SNe Ia+OHD data, with very similar values of $\chi^{2}_{min}$, again due to the similarity of the behavior of the two models, especially at late times. 

A remarkable result is that the viscous $\Lambda$WDM model exhibits a slightly lower value of $\chi_{min}^{2}$ for the SNe Ia+OHD data than the $\Lambda$CDM model, despite the fact that the $\Lambda$WDM model has a greater value of BIC than the $\Lambda$CDM model. This translate into a better fit for the $\Lambda$WDM model because the two extra free parameters of the viscous $\Lambda$WDM model are a consequence of a more general description of the DM component, which takes into account a warm nature and a non-perfect fluid description, suggest by previous investigations that use this alternatives to face tensions of the standard model, and they are not been added by hand in order to force a better fit. Even more, the $\Lambda$CDM model assumes beforehand a CDM (one less free parameter) and all their matter components are describes as perfect fluids (another less free parameter) which leads to a good fit of the combined SNe Ia+OHD data with less free parameters but, the price to pay is the problems mentioned above that the $\Lambda$CDM model experience today. Therefore, we have an alternative model which goes beyond to the standard $\Lambda$CDM model by considering a dissipative WDM, which has the capability to describe the SNe Ia and OHD data in the same way as the $\Lambda$CDM model, together with a more general description of DM nature.

By the other hand, the best fit values contrasted with the combined SNe Ia + OHD data, for $\gamma$ and $\Omega_{\xi_{0}}$ are given by $1.02555^{+0.04453}_{-0.02424}$ and $0.01633^{+0.04185}_{-0.01627}$ respectively at $3\sigma$ CL. Note that, both deviations around the mean value satisfy the near equilibrium condition  Eq. \eqref{actualnearcondition}, which means, that these values are compatible with our previous theoretical thermodynamics conclusion ($\Omega_{\xi_{0}}<1$ and $\gamma\neq1$). For the small values of $\Omega_{\xi_{0}}$ and $\gamma$ given by the data, we are far from the near equilibrium condition at $H=21.61855H_{0}$ according to Eq. \eqref{lhm1} , which means that, we are far from near equilibrium at a redshift of $z=11.63228$ according to Eq. \eqref{Hconstraint}, by then, we can ensure the near equilibrium condition for the actual data measurement at $z\sim2.3$.

Also, is important to mention that, with the data measurement we can obtain the actual size of dissipation for our $\Lambda$WDM model. Note that, according to our dimensionless expression, $\Omega_{\xi_{0}}=3\xi_{0}H_{0}$, $\xi_{0}$ has dimension of time, and the bulk viscosity is given by $\xi=\xi_{0}\rho$, 
which writing in dimensionless full ($c\neq1$), is given by $\xi=\xi_{0}\varepsilon$, being $\varepsilon$ the energy density of matter (that has the same dimension of pressure according to the EoS), then $\xi$ would have the viscous units of $Pa\times s$. Follow the cosmological constraint, the maximum value for $\Omega_{\xi_{0}}$ leads to consider $\Omega_{\xi_{0}}<0.05818$ (a generalization of our theoretical constraint $\Omega_{\xi_{0}}<1$ ), by then, using the definition of $\Omega_{\xi_{0}}$ we have the follow constraint on bulk viscosity 
\begin{equation}
  \xi<\varepsilon\times\frac{0.05818}{3H_{0}},  
\end{equation}
if we introduce the critical density given by $\varepsilon_{0}=\frac{3c^{2}H^{2}_{0}}{8\pi G}$ and since our best fit values of Hubble parameter $H_{0}$ and matter density $\Omega_{m0}$ at $3\sigma$ level are $H_{0}=70.9238^{+0.02584}_{-0.02680}\frac{km/s}{Mpc}$ 
and $\Omega_{m}=0.26073^{+0.04741}_{-0.04073}$ respectively, then, if we consider for instance the means values for $H_{0}=70.9238\frac{km/s}{Mpc}=\left(4.3507\times 10^{17}s\right)^{-1}$ 
and $\Omega_{m0}=0.26073$, the restriction over the upper limit of bulk viscosity would be 
\begin{equation}
  \xi<1.87082\times 10^6 Pa\times s,  
\end{equation}
for this particular case of values, if we consider the deviation around the mean values the upper limits remains of the order of $\xi\lesssim 10^6 Pa\times s$. Note that, from theoretical ground ($\Omega_{\xi_{0}}<1$) and using the value of $H_{0}$ and $\Omega_{m}$ given by \cite{Riess:2019cxk,Planck2018} we will find that 
\begin{equation}
    \xi_{0}<3.628\times10^{7} Pa\times s,
\end{equation}
 a similar value are found by B. D. Normann and I. Brevik in \cite{valorxi1}, where they consider a model with a viscous DM component with bulk viscosity of the form $\xi=\xi_{0}\rho^{\frac{1}{2}}$, and DE component given by the CC, they also suggest (see also \cite{valorxi2}) that a value $\xi_{0}\sim10^{6}Pa\times s$ for the present viscosity is reasonable. Also, other investigations suggest the same orders of magnitude \cite{valorxi1,xi3,produccionentropia5,valorxi2}, but since these models are not identical, it is expect some discrepancies with our results.

\section{Conclusions}\label{seccionfinal}
We have discussed throughout this work the near equilibrium condition, entropy production, and cosmological constraint of a  cosmological model filled with a dissipative WDM, where the bulk viscosity is proportional to the energy density, and a positive CC, which is described by an exact solution previously found in \cite{primerarticulo}. Assuming the condition given by Eq. \eqref{condicionBigRip} this solution behaves very similar to the standard cosmological model for small values of $\Omega_{\xi_{0}}$ as we can see in Fig. \ref{figuraem1}, avoiding the appearance of a future singularity in a  finite time (Big-Rip).

We have shown that the presence of the CC together with a small viscosity from the expression \eqref{lhm1}, and considering the condition given by the Eq. \eqref{condicionBigRip}, leads to a near equilibrium regime for the WDM component. 

The WDM component has a temperature which decreases very slowly with the cosmic expansion as a result of being close to the near equilibrium condition, contrary to the non physically  behavior found  for the dust case ($\gamma=1$), where the temperature increase with the scale factor. Besides, we have shown that the second law of thermodynamics is fulfilled as long we satisfied the conditions \eqref{condicionBigRip} and \eqref{gammaWDM} (near equilibrium condition). On the contrary, the entropy production would diverge in a finite time if a Big-Rip singularity occurs.

To fulfill the two criteria discussed in this work, we need to have then $\Omega_{\xi_{0}}<1$ and a WDM component with a EoS satisfying the constraint given in Eq.\eqref{actualnearcondition}. 

It is important to mention that in our WDM model we need to have $\Omega_{\xi_{0}}\ll\gamma-1$. For small values of $\Omega_{\xi_{0}}$ (in particularly $\Omega_{\xi_{0}}=0.001$), the model enter into agreement with some previous results found, for example in \cite{wefectivo}, where cosmological bounds on the EoS for the DM were found, and the inclusion of the CC is considered. The bounds for a constant EoS are  $-1.50\times10^{-6}<\omega<1.13\times10^{-6}$ ($\omega=\gamma-1$) if there is no entropy production and $-8.78\times10^{-3}<\omega<1.86\times10^{-3}$ if the adiabatic speed of sound vanishes, both at $3\sigma$ of confidence level. Another example can be found in \cite{wefectivo2} where, using WMAP+BAO+HO observations, the EoS at the present time is given by $\omega=0.00067^{+0.00011}_{-0.00067}$.

Highlighting again, we have to mention that the asymptotic behavior of the exact solution at the infinite future, given by \eqref{HdeSitter}, corresponds to the usual de Sitter solution, which indicates that this solution describes a dissipative WDM that could reproduce the same asymptotic behavior of the standard model. As long as the exact solution tends to this value, the near equilibrium condition \eqref{nearequilibrium} can be satisfied. Of course, the de Sitter solution has a constant temperature, according to Eq. \eqref{temperatura} and, since in this case we have a null density and null pressure of the fluid, the entropy production is zero, according to Eq. \eqref{entropyproduction} 

We have shown in this work that, previously to any constraining from the cosmological data, the study of thermodynamics consistences required by the Eckart approach, such as the near equilibrium condition and entropy production, leads to important constraints on the cosmological parameter, such as the given one by Eq. \eqref{actualnearcondition}, which implies the necessity of WDM, in agreement with some previous results found in \cite{wefectivo,wefectivo2,wefectivo3}. On the other hand, the constrain \eqref{condicionBigRip} tells us that Big-Rip singularities are avoided at late times, if the near equilibrium condition is preserved. Even though, the exact solution explored behaves very similar to the standard model, and open the possibility of a more realistic fluid description of the DM containing dissipation processes, within the Eckart's framework,  giving us physically important clues about the EoS of this component ($\gamma$) and the size of dissipation involved ($\Omega_{\xi_{0}}$). In this sense, this thermodynamics analysis had the aim of finding a model that satisfies the near equilibrium condition, together with a proper behavior of the temperature and entropy production, helping us to find the best prior definition for the cosmological constraints discussed in Sec.\ref{constraints} .

The result of the cosmological constraint are shown in Table \ref{tab:bestfits}, as well as their corresponding goodness of fit criteria and the joint analysis for the SNe Ia and OHD data. The more greater value of BIC for the $\Lambda$WDM in comparison to the value obtained for the $\Lambda$CDM model is a reflection of considering a more general theory with two more free parameters ($\gamma$ and $\Omega_{\xi_{0}}$), than the standard $\Lambda$CDM model ($\gamma$ fixed to $1$ and a perfect DM component), obtaining also a slightly better fit for the combined SNe Ia + OHD data for the $\Lambda$WDM model, being in this sense a more realistic model that describes in the same way (slightly better) the SNe Ia and OHD data like the standard cosmological model does.

With the cosmological values $\Omega_{\xi_{0}}$ and $\gamma$ at $3\sigma$ CL obtained with the combined SNe Ia + OHD data, we can conclude that the near equilibrium condition is full satisfies even for a redshift of $z=11$ according to Eq. \eqref{Hconstraint}, and then, and proper physical behavior of the temperature (decreasing function with the scale factor), and entropy production (without future Big-Rip singularity), is obtained. Even more, our data analysis suggest that the actual value of  bulk viscosity has an upper limit of the order of $10^{6}Pa\times s$ in agreement with some previous investigation. 
It is also important to mention that, the model theoretically (and even from the cosmological data) does not rule out values smaller than this, since the space of values obtained from the theoretical ground at the present time is given by $0<\Omega_{\xi_{0}}<1$, and allows in principle, get smaller values, that can be more acceptable from the viscous hydrodynamic point of view.


Finally, the main contribution of the present study is to show that the standard model with the two extensions suggest, i.e., a dissipative WDM component, in order to face the recent cosmological tension found, fulfill the criteria of a consistent relativistic fluid description under some constraints, which also describe the combined SNe Ia + OHD data in the same way as the $\Lambda$CDM. Therefore, our exact solution describes a physically viable model for a dissipative WDM component, which is supported by many investigations that have extended the possibilities for the DM nature to face tensions of the standard model.

\section*{Acknowledgments}
Norman Cruz acknowledges the support of Universidad de Santiago de Chile (USACH), through Proyecto DICYT N$^{\circ}$ 042131CM, Vicerrector\'ia de Investigaci\'on, Desarrollo e Innovaci\'on. Esteban Gonz\'alez acknowledges the support of Direcci\'on de Investigaci\'on y Postgrado at Universidad de Aconcagua. Jose Jovel acknowledges ANID-PFCHA/Doctorado Nacional/2018-21181327.

\bibliography{bibliografia.bib}
\end{document}